\documentstyle[11pt,newpasp,twoside,epsf]{article}
\def\edcomment#1{\iffalse\marginpar{\raggedright\sl#1\/}\else\relax\fi}
\reversemarginpar

\def\hii{H\thinspace{$\scriptstyle{\rm II}$}~}

\def\eg{{\it e.g.},~}
\def\3he{$^3$He}
\def\4he{$^4$He}
\def\6li{$^6$Li}
\def\7li{$^7$Li}
\def\3h{$^3$H}
\def\Yp{Y$_{\rm P}$~}

\begin{document}
\title{Primordial Nucleosynthesis For The New Millennium}
\author{G. Steigman}
\affil{Departments of Physics and Astronomy; The Ohio State University; 
174 West 18th Avenue; Columbus, OH 43210 USA}

\begin{abstract}
The physics of the standard hot big bang cosmology ensures that the early 
Universe was a primordial nuclear reactor, synthesizing the light nuclides 
(D, \3he, \4he, and \7li) in the first 20 minutes of its evolution.  After 
an overview of nucleosynthesis in the standard model (SBBN), the primordial 
abundance yields will be presented, followed by a status report (intended 
to stimulate further discussion during this symposium) on the progress along 
the road from observational data to inferred primordial abundances.  Theory 
will be confronted with observations to assess the consistency of SBBN and 
to constrain cosmology and particle physics.  Some of the issues/problems 
key to SBBN in the new millenium will be highlighted, along with a wish 
list to challenge theorists and observers alike.
\end{abstract}

\section{Introduction}

Among the quantitative, ``hard" sciences, astronomy has traditionally 
been scorned, with particular disdain reserved for cosmology.  No more.  
In the decade of the nineties the combination of an avalanche of high 
quality observational data and theoretical advances driven by enhaced 
computer (and brain) power, have succeeded in transforming cosmology to 
a precise science.  In this introductory lecture to IAU Symposium 198 
on The Light Elements and Their Evolution it is my intent to describe 
primordial nucleosynthesis in this precision era of cosmology and to 
highlight the challenges, along with some goals, for the new millennium.  
After a brief review of the important physics during the era of primordial 
nucleosynthesis in the standard, hot big bang cosmological model (SBBN), 
I will present an overview of the predicted primordial abundances, 
emphasizing the generally very small theoretical uncertainties.  These 
will then be compared to the present best estimates (including their 
uncertainties) of the primordial abundances inferred from current 
observational data.  After assessing the consistency of SBBN, I will 
explore what SBBN has to offer to Cosmology and to Particle Physics 
and, what Cosmology may teach us about SBBN.  I will conclude with a 
summary of the key issues/problems confronting SBBN and with a wish 
list of topics I hope will be addressed during this meeting -- and beyond.

\section{An Early Universe Chronology}

Our story begins when the Universe is a few tenths of a second old and 
the temperature of the cosmic background radiation has dropped to a few 
MeV as the Universe expanded and cooled from its denser, hotter infancy.  
At this time (and earlier) the density and average energy of colliding 
particles is so high that even the weak interactions occur sufficiently 
rapidly to establish equilibrium.  In particular, at this stage all flavors 
of neutrinos ($e, \mu, \tau$) are in thermal equilibrium with the cosmic 
background radiation (CBR) photons and with the copius electron-position 
pairs present ($\nu_{i} + \bar{\nu}_{i} \leftrightarrow e^{+} + e^{-} 
\leftrightarrow \gamma + \gamma$).  However, as the Universe ages beyond 
a few tenths of a second and the temperature drops below a few MeV, these 
weak interactions become too slow to keep pace with the rapid expansion of 
the Universe and the neutrinos decouple from the CBR.  The electron-type 
neutrinos continue to play a role in transforming neutrons into protons 
and, vice-versa ($p + e^{-} \leftrightarrow n + \nu_{e}$, $n + e^{+} 
\leftrightarrow p + \bar{\nu}_{e}$, $n \leftrightarrow p + e^{-} + \bar
{\nu}_{e}$).  As the temperature continues to drop, less massive protons 
are favored over the more massive neutrons and the $n/p$ ratio falls (roughly 
as $e^{-\Delta m/kT}$, where $\Delta m$ is the neutron -- proton mass 
difference $\sim 1.3$~MeV).  After the temperature drops below 800~keV 
or so, when the Universe is a few seconds old, even these weak interactions 
become too slow to keep pace with the expansion and the neutron-to-proton 
ratio ``freezes out" (in fact, the ratio continues to decrease, albeit 
very slowly).  All the while, neutrons and protons have been colliding, 
occasionally forming deuterons ($p + n \rightarrow D + \gamma$).  However, 
the deuterons find themselves bathed in a high density background of 
energetic CBR photons which quickly photodissociate them ($D + \gamma 
\rightarrow p + n$) before they can find a proton or neutron and form 
the more tightly bound, less fragile, \3h or \3he nuclei.  Since, as we 
shall see, there are roughly nine to ten orders of magnitude more CBR 
photons than nucleons in the Universe, the deuteron ``stepping-stone" to 
further nucleosynthesis is absent until the temperature drops sufficiently 
low so that even in the high-energy tail of the black-body spectrum there 
are too few photons to prevent the deuteron from acting as a catalyst for 
primordial nucleosynthesis.  This critical temperature, which is weakly 
(logarithmically) dependent on the nucleon abundance (the nucleon-to-photon 
ratio $\eta$), is roughly 80~keV.  Now, at last, when the Universe is a 
few minutes old, Big Bang Nucleosynthesis finally commences.  However, 
the Universe was a fatally flawed nuclear reactor, cooling and diluting 
rapidly as it aged.  When the Universe is some 10 -- 20 minutes old 
($\sim 1000$~sec) and the temperature has dropped below 30~keV or so, 
the coulomb barriers preventing nuclear reactions between charged nuclei 
and protons and among charged nuclei become insurmountable (in the short 
amount of time available) and primordial nucleosynthesis comes to an 
abrupt end.  In this all too brief but shining era there has been time 
to synthesize (in abundances comparable to those observed or observable) 
only the lightest nuclides: D, \3he, \4he, and \7li.  In ``standard" (a 
homogeneous Universe, expanding isotropically with the particle content 
of the standard model of particle physics in which there are three flavors 
of light ($m \ll $~MeV) or massless neutrinos) big bang nucleosynthesis 
(SBBN) the abundances (relative to protons $\equiv$~ hydrogen) of these 
four nuclides are determined by only one free parameter, the present epoch 
nucleon-to-photon ratio $\eta$ ($\eta \equiv (n_{\rm N}/n_{\gamma})_{0}, 
~\eta_{10} \equiv 10^{10}\eta$).

\section{SBBN-Predicted Primordial Abundances}

Once the deuterium photodissociation bottleneck is breached primordial 
nucleosynthesis begins in earnest, quickly burning D to \3h, \3he and 
\4he.  The higher the nucleon density, the faster D is destroyed.  The 
same is true of \3h (which, if it survives will decay to \3he) and \3he.  
Thus, the primordial abundances of D and \3he are determined by the 
competition between the nuclear reaction rates and the universal expansion 
rate.  The former rate depends on the overall density of the reactants -- 
the nucleon density.  Since all densities decrease as the Universe expands, 
it is convenient to quantify the nucleon density by specifying the {\it 
ratio} of the nucleon density to the photon density (measured after $e^{+}
e^{-}$ annihilation which enhances the Universe's photon budget) $\eta$.  
Since observations of the cosmic background radiation (CBR) temperature 
(T = 2.73~K) determine the present density of CBR photons, a knowledge 
of $\eta$ is equivalent to a determination of the present mass density 
in nucleons (``baryons" $\equiv $~B).  In terms of the density parameter
$\Omega_{\rm B}$ (the ratio of the mass density to the critical mass density)
and the present value of the Hubble parameter (H$_{0} \equiv 100h$~km/s/Mpc),
$\eta_{10} = 273\Omega_{\rm B}h^{2}$.  As $\eta$ increases the surviving 
abundances of D and \3he decrease; since the \3he nucleus is more tightly 
bound than the deuteron, the decrease of the \3he/H ratio with $\eta$ is 
less rapid than that of D/H.  

In contrast to D and \3he, the primordial abundance of \4he is not reaction 
rate limited since the nuclear reactions building helium-4 are so rapid 
that virtually all neutrons available when BBN commences are incorporated 
into \4he.  As a result the \4he abundance, conventionally presented as 
the mass fraction of all nucleons which are in \4he, Y$_{\rm P}$, is {\it 
neutron limited}.  Since the neutron-to-proton ratio is determined by the 
competition between the (charged-current) weak interactions which mediate 
the transformation of neutrons into protons (and, vice-versa) and the 
universal expansion rate, \Yp is sensitive to the universal expansion 
rate at the time the $n/p$ ratio ``freezes" and when the deuterium 
photodissociation barrier disappears.  Since the universal expansion rate 
is controlled by the total energy density, \Yp provides an important test 
of cosmology and of particle physics in the early Universe (Steigman, 
Schramm \& Gunn 1977).  It should be noted that \Yp is not entirely 
insensitive to the nucleon density since the higher $\eta$, the earlier 
the photodissociation barrier is overcome.  At earlier times when the 
temperature is higher, fewer neutrons have been transformed into protons 
and are available for incorporation into \4he.  As a result, \Yp increases 
logarithmically with $\eta$.

There is no stable nucleus at mass-5 and this presents a gap in the 
road to the synthesis of nuclei heavier than \4he.  In order to bridge 
the gap nuclear reactions must occur among nuclei with two or more 
nucleons.  But, the abundances of D, \3h, and \3he are small and the 
coulomb barriers (especially between \3he and \4he and between \4he 
and \4he) suppress these reactions as the Universe expands and cools.  
As a result, there is very little ``leakage" to nuclei beyond mass-4; 
as a corollary, virtually all the \4he formed, survives.  The only 
heavier nucleus produced primordially in an abundance comparable to 
that observed (or, even, observable with current technology) is \7li, 
whose BBN abundance is some 4 -- 5 orders of magnitude smaller than 
that of D and \3he.  The absence of a stable nucleus at mass-8 provides 
another gap preventing the production of astrophysically interesting 
abundances of any heavier nuclei.

As will become clear in our subsequent discussion, the ``interesting" 
range of $\eta$ is $\eta_{10} = 1 - 10$ ($\Omega_{\rm B}h^{2} = 0.004 
- 0.037$), so we focus our discussion here on values of $\eta$ in this 
range.  In the current precision era of BBN {\it most} of the nuclear
reactions relevant to the synthesis of the light elements have been
measured to reasonable accuracy at energies directly comparable to the
thermal energies at the time of primordial nucleosynthesis (\eg see 
Nollett, this volume).  As a result, the theoretical uncertainties in
the BBN-predicted abundances are generally quite small.  For $\eta$ in
the above range, the 1$\sigma$ uncertainties in D/H and \3he/H vary
from 8 -- 10\%.  Since \4he is most sensitive to the very well measured
weak interaction rates, the error in SBBN-predicted \Yp is very small 
(0.2 -- 0.5\% or, $\sigma_{\rm Y}$ = 0.0005 -- 0.0011).  In contrast,
larger uncertainties, of order 12 -- 21\%, afflict the predicted 
primordial abundance of \7li.  

Since this Symposium devoted much discussion to \7li, and space-limitations
here prevent me from discussing all the light elements in detail, I will
concentrate in the following on the two key light elements, deuterium and 
helium-4.  In Figure 1 is shown the relation between the BBN-predicted
abundances of D and \4he.  The band going from upper left to lower right
represents the $\pm 2\sigma$ range of uncertainties in the primordial 
abundances ((D/H)$_{\rm P}$ and Y$_{\rm P}$).  Low D/H (high $\eta$) 
corresponds to high \Yp and high D/H (low $\eta$) corresponds to low 
Y$_{\rm P}$.  This anticorrelation will be very important when we 
confront the predictions of SBBN with the observational data.

\section{Precise (Accurate?) Primordial Abundances}

To test SBBN and fully exploit the opportunities it offers to constrain
cosmology (e.g., the baryon density) and particle physics (e.g., new
particles with weak or weaker-than-weak interactions) requires that 
observational data be used to pin down the primordial abundances of the
light elements to precisions as good as (or, better than) those of the
SBBN predictions.  As we approach the new millennium there is good news
along with some bad news.  The good news is that new detectors on ever
larger telescopes which cover the spectrum from radio to x-ray energies
and beyond are providing very high quality data, leading to inferred
abundances of high statistical accuracy.  Furthermore, the abundances 
of the light elements are determined from observations which differ 
from element to element in the telescopes and techniques employed as 
well as in the astrophysical sites explored.  As a result, insidious 
correlated errors between and among the various element abundances are 
unlikely to be a problem.  The good news is also responsible for the 
bad news.  Since the statistical errors have become so small, systematic 
errors now tend to dominate the uncertainties in the derived primordial 
abundances.  As Bob Rood has said during this Symposium, estimating 
systematic errors is an oxymoron.  When a potential source of systematic 
error is identified, observations can (and should) be designed to 
eliminate or bound its contribution to the error budget.  It is a 
pointless and potentially misleading exercise to ``estimate" the magnitude 
of unidentified systematic errors.  In part to remind us that our precise 
abundance determinations may not be accurate, and in part to challenge 
our observational colleagues who have done such a magnificent job of 
reducing the statistical errors, I will try to focus on the potential 
sources of systematic uncertainty (when I can identify them) in the 
following overview of the current observational status.

\subsection{Deuterium}

As J. Linsky (this volume) has reminded us, the deuterium abundance 
in the local interstellar medium (the local interstellar cloud: LIC) 
is known very accurately: (D/H)$_{\rm LIC} = 1.5 \pm 0.1 \times 
10^{-5}$ (Linsky 1998).  Since deuterium is only destroyed during 
the evolution of the Galaxy (Epstein, Lattimer \& Schramm 1976), the 
LIC abundance provides a lower bound to its primordial (pre-Galactic) 
value.  This bound is strong enough to bound the nucleon density from 
above ($\eta_{10} \la 10$; $\Omega_{\rm B}h^{2} \la 0.04$), ensuring 
that baryons cannot ``close" the Universe ($\Omega_{\rm B} \ll 1$), 
nor even dominate its present mass density ($\Omega_{\rm B} \ll 
\Omega_{\rm M} \approx 0.3 - 0.4$).  Thus, local observations of 
deuterium, combined with the {\it assumption} of the correctness 
of SBBN (which we must test), already reaps great rewards: the 
mass-energy density of the Universe must be dominated by unseen 
(``dark") non-baryonic matter.  To go beyond (in the quest for the 
primordial deuterium abundance) we must look for observing targets 
which are less evolved than the LIC.  The presolar nebula is one such 
site.  From solar system observations of \3he reported by G. Gloeckler 
(this volume), it is possible to infer the presolar deuterium abundance 
(Geiss \& Reeves 1972; Geiss \& Gloecker 1998): (D/H)$_{\odot} = 1.9 
\pm 0.5 \times 10^{-5}$.  Although marginally higher than the LIC 
abundance, the larger errors prevent us from using this determination 
to improve on our previous bounds from the LIC.  What this result does 
indicate is that there has been very little (if any) evolution in the 
D-abundance in the solar vicinity of the Galaxy in the last 4.5~Gyr.  
This is consistent with a large class of Galactic chemical evolution 
models discussed by M. Tosi (this volume) which point to only a modest 
overall destruction of primordial deuterium by a factor of 2 -- 3 
(Tosi et al. 1998).  If this theoretical estimate is combined with the 
LIC abundance, we may estimate the primordial abundance: (D/H)$_{\rm P} 
\approx 2.6 - 5.1 \times 10^{-5}$ ($\sim 2\sigma$).  Although possibly 
model dependent, this estimate is in remarkable agreement with the 
2 -- 3 determinations of D/H in high-redshift, low-metallicity (hence 
very nearly primordial) Ly-$\alpha$ absorbers illuminated by background 
QSOs described by D. Tytler and S. Levshakov (this volume).  The 
data and analysis of Burles \& Tytler (1998a,b: BT) suggests that 
(D/H)$_{\rm P} = 2.9 - 4.0 \times 10^{-5}$ ($\sim 2\sigma$).  Notice 
that the 1$\sigma$ uncertainty in the observationally determined 
primordial abundance, $\sim 8$\%, is impedance-matched to the 
$\sim 8$\% SBBN theoretical uncertainty cited earlier.  However, 
lest we risk dislocating a shoulder while patting ourselves on the 
back at the triumph of such wonderful data, we should not ignore 
the claim (Webb et al. 1997; Tytler et al. 1999) that the deuterium 
abundance in at least one Ly-$\alpha$ absorption system may be much 
higher.  This is a reminder that while any determination of the 
deuterium abundance anywhere in the Universe (LIC, solar system, 
Ly-$\alpha$ absorbers, etc.) provides a {\it lower} bound to primordial 
deuterium, finding an upper bound is more problematic.  Indeed, in 
some absorbing systems it may be impossible to distinguish D-absorption 
from that due to hydrogen in an interloping, low column density, 
``wrong-velocity" system.  Thus, the deuterium abundance inferred 
from absorption-line data may only provide an {\it upper} bound to 
the true deuterium abundance.  Since the low-Z, high-z QSO absorbing 
systems hold the greatest promise of revealing for us nearly unevolved, 
nearly primordial material, we look forward to the time when we can 
use the {\it distribution} of D/H values from more than a handful of 
such systems to eliminate -- statistically -- the uninvited contribution 
to the inferred primordial deuterium abundance from such interlopers.  
Keeping this in mind, in the following I will, nevertheless, use the 
BT determination when confronting theory with data. 

\subsection{Helium-4}

In contrast to deuterium whose primordial abundance only decreases 
as pristine gas is incorporated into stars, stars burn hydrogen to 
helium.  As a result, the \4he observed anywhere in the Universe is 
an unknown mixture of primordial and stellar-produced helium.  It has 
long been appreciated that to minimize the uncertain correction due 
to the debris of stellar evolution, it is best to concentrate on \4he 
abundance determinations in the lowest-metallicity regions available.  
These are the low-Z, extragalactic \hii regions which have been discussed 
by K. Olive, T. Thuan, and S. M. Viegas at this Symposium (this volume).  
The reader is urged to consult their papers for details; here I will 
merely summarize my view of the current status of the determination 
of the primordial \4he mass fraction Y$_{\rm P}$.  Several years ago 
Olive \& Steigman (1995: OS) gathered together the data from the literature 
(dominated by the data assembled by Pagel et al. 1992).  More recently 
Olive, Skillman \& Steigman (1997: OSS) supplemented this with newer 
data (some of it, unfortunately, still unpublished).  Using a variety 
of approaches such as the regression of Y on the oxygen and/or nitrogen
abundances and the weighted means of Y in the lowest metal-abundance 
\hii regions, OSS concluded that \Yp = 0.234 $\pm$ 0.003 (note that, 
in contrast to the published (OSS) result, this value is obtained when 
the NW region of IZw18, suspected of being contaminated by underlying 
stellar absorption, is excluded from the fit, and the newer data of 
Izotov, Thuan and collaborators is not included).  Izotov, Thuan 
and their collaborators (Izotov, Thuan, \& Lipovetsky 1994, 1997; 
Izotov \& Thuan 1998(IT); Thuan, this volume) have been systematically 
observing a mostly independent set of \hii regions.  Although, as 
with the data employed in the OS and OSS studies, they ignore the 
ionization correction ($icf \equiv 1$), they take special care with 
the correction for collisional excitation.  IT (also Thuan, this volume) 
find Y$_{\rm P}({\rm IT}) =  0.244~\pm$ 0.002.  Comparing the IT and OSS 
estimates of \Yp we find that difference between the two \Yp estimates 
far exceeds the statistical errors, suggesting systematic differences 
in the acquisition and/or analysis of the data samples.  In a recent 
discussion which attempted to account for these unidentified systematic 
differences, Olive, Steigman \& Walker (1999: OSW) combined the 2$\sigma$ 
ranges for each determination to conclude: \Yp = 0.238 $\pm$ 0.005; at 
the 2$\sigma$ level, \Yp $\leq 0.248$.  Note, that this is also the 
2$\sigma$ upper bound to the IT data alone.  Since, as we shall see 
shortly, it is the upper bound which is crucial to testing the consistency 
of SBBN, in the following we shall adopt the IT value (and error estimate) 
for the primordial abundance of \4he.  

Recently, Viegas, Gruenwald \& Steigman (1999: VGS; see Viegas \& 
Gruenwald, this volume) have emphasized the importance of the ionization 
correction which has heretofore been ignored.  VGS suggest that the 
IT helium abundance (Y$_{\rm P}$) should be reduced by 0.003 to account 
for unseen neutral hydrogen in regions where the helium is still ionized 
in \hii regions ionized by young, hot, metal-poor stars.  In subsequent 
comparisons I shall explore the implications of adopting Y$_{\rm P}({\rm 
VGS}) = 0.241~\pm$ 0.002.

\subsection{Helium-3 and Lithium-7}

The cosmic history of the two other light nuclides produced in
astrophysically interesting abundances during SBBN, \3he and 
\7li, is considerably more complex than that of D or \4he, which 
limits their utility as probes of the consistency of SBBN.  \3he 
is destroyed in the hotter interiors of all stars, but some \3he 
does survive in the cooler, outer layers.  For lower mass stars 
this \3he survival layer increases and, indeed, newly synthesized 
\3he is produced by incomplete hydrogen burning.  The competition 
between destruction, survival, and synthesis complicates the Galactic
history of the \3he abundance.  Nonetheless, since any deuterium 
incorporated into stars is first burned to \3he, the apparent lack 
of enhanced \3he (see Bania \& Rood, this volume) argues against 
a very large pre-Galactic abundance of deuterium (Steigman \& Tosi 
1995).  For further discussion of the evolution of \3he see Bania 
\& Rood (this volume).

As with \3he, any \7li incorporated into stars is quickly burned 
away.  However, fusion and spallation reactions between cosmic ray 
nuclei and those in the interstellar medium are a potent source of 
\7li (as well as of \6li, $^7$Be, $^{10}$B, and $^{11}$B).  It is
also likely that there are stellar sources of \7li as indicated 
by the sample of lithium-rich red giants (V. Smith, this volume).  
Since the abundance of lithium in the solar system and in the 
interstellar medium (``here and now") greatly exceeds that in the 
very metal-poor halo stars (T. Beers \& S. Ryan, this volume), the 
latter likely provide the closest approach to a nearly primordial 
sample.  Since a significant fraction of this Symposium is devoted 
to lithium, I will defer here to those other discussions except to 
comment that, within the theoretical and observational uncertainties, 
the primordial abundances inferred from the observational data are 
consistent with SBBN constrained by the confrontation with D and \4he.

\section{Confrontation Of SBBN With Data}

Although SBBN does lead to the prediction of the abundances of D, 
\3he, \4he, and \7li, the currently best-constrained primordial 
abundances are those of deuterium and helium-4 which we are 
concentrating on in this status report.  For each value of $\eta$, 
SBBN predicts a pair of (D/H)$_{\rm P}$ and \Yp values.  Therefore, 
in SBBN there is a unique connection between (D/H)$_{\rm P}$ and 
\Yp which, allowing for the theoretical uncertainties discussed 
above, is shown as the band (solid lines) in Figure 1 going from 
the upper left to the lower right (2$\sigma$ uncertainties).  Note 
that high-helium correlates with low-deuterium and, vice-versa.  
Also shown as the dotted ellipse in Figure 1 is the contour of the 
(independent) 2$\sigma$ uncertainties in the BT deuterium abundance 
and the IT helium-4 mass fraction.   

\begin{figure}
\vspace{1.in}
\plotfiddle{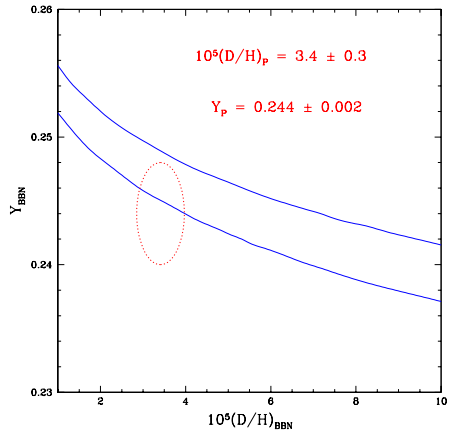}{1in}{0}{30}{30}{-100}{-50}
\caption{The SBBN-predicted \4he mass fraction \Yp as a function 
of the SBBN-predicted primordial deuterium-to-hydrogen ratio D/H 
is shown (at the $\pm 2\sigma$ level) by the solid lines.  The 
dotted ellipse is the 95\% contour of the BT deuterium abundance 
and the IT \4he mass fraction (see the text).}
\end{figure}

Although the overlap between theory and data is not complete, 
Figure 1 shows that, at the $\sim 2\sigma$ level, the predictions 
of SBBN are consistent with current observational data.  This is 
a dramatic success for the standard hot, big bang cosmological 
model.  Of course it is not at all surprising that some value 
of $\eta$ may be found to provide consistency with the inferred 
primordial deuterium abundance.  But there was no guarantee at 
all that the helium-4 abundance corresponding to this choice would 
bear any relation to its inferred primordial value.  Consistency 
with the BT D-abundance limits the nucleon abundance to the range 
(2$\sigma$) $\eta_{10} = 4.4 - 5.9$ or, $\Omega_{\rm B}h^{2} = 
0.016 - 0.022$.  For $\eta$ in this range there is consistency, 
within the theoretical and observational uncertainties, between 
the SBBN-predicted and observationally inferred primordial abundances 
of \3he and \7li as well.  Four for the price of one!  There is, 
of course, one more test -- and opportunity -- offered by this 
result.  This SBBN-inferred nucleon abundance must also be 
consistent with present epoch estimates of the baryon density.  
Indeed, the SBBN-determined value of $\Omega_{\rm B}$ is {\it 
larger} than estimates (Persic \& Salucci 1992) of the ``luminous" 
matter in the Universe suggesting that the majority of baryons 
are ``dark".  This is good ($\Omega_{\rm B} > \Omega_{\rm LUM}$); 
the opposite would have been a disaster.  This early-Universe 
estimate of the baryon density is in good agreement with that 
inferred from the X-ray cluster baryon fraction (Steigman, Hata 
\& Felten 1999) and with the independent estimate from the 
Ly-$\alpha$ forest (Weinberg et al. 1997) discussed below.

\subsection{What BBN May Do For Cosmology}

X-ray clusters likely provide a ``fair" sample of the universal
baryon {\it fraction} $f_{\rm B}$ (White et al. 1993; Steigman 
\& Felten 1995; Evrard, Metzler, \& Navarro 1996) which, when 
combined with the SBBN-inferred baryon density $\Omega_{\rm B}$, 
leads to a ``clean" prediction, independent of detailed cosmological 
models, of the overall matter density $\Omega_{\rm M}$.  If the 
results presented here are combined with the determination of 
$f_{\rm B}$ from Evrard (1997), and with a Hubble parameter $h 
= 0.70 \pm 0.07$ (Mould et al. 1999), we predict $\Omega_{\rm M} 
= 0.35 \pm 0.08$, in excellent agreement with several other recent, 
independent determinations.  For example, a lower bound to the cosmic 
baryon density follows from the requirement that the high-redshift 
intergalactic medium contain enough neutral hydrogen to produce
the Ly-$\alpha$ absorption observed in quasar spectra.  According
to Weinberg et al. (1997), depending on estimates of the quasar
UV background intensity, this lower bound corresponds to $\eta_{10}
\ga 3.4 - 4.9$, in excellent agreement with the SBBN prediction
based on the BT deuterium determination.  Note that this lower
bound from the Ly-$\alpha$ absorption forbids (in the context of
SBBN) the primordial deuterium abundance to be any larger than
$\sim 8 \times 10^{-5}$, largely excluding the one surviving
claim of high D (Webb et al. 1997).

Indeed, if the SBBN results are combined with the magnitude-redshift 
data from surveys of high-redshift Type Ia supernovae (Garnavich et 
al. 1998; Perlmutter et al. 1999) which bound a linear combination 
of $\Omega_{\rm M}$ and the cosmological constant $\Omega_{\Lambda} 
\equiv \Lambda/3$H$_{0}^{2}$, we may also constrain the cosmological 
constant ($\Omega_{\Lambda} = 0.80 \pm 0.20$), the curvature 
($\Omega_{k} \equiv 1 - (\Omega_{\rm M} + \Omega_{\Lambda}) = 
-0.15 \pm 0.25$), and the deceleration parameter ($q_{0} = 
\Omega_{\rm M}/2 - \Omega_{\Lambda} = -0.62 \pm 0.18$).

\subsection{What Cosmology May Do For BBN}

As we have just seen, the SBBN-determined baryon density is 
consistent with that determined or constrained by observations 
of the Universe during its present or recent evolution.  We 
may turn the argument around and ask what baryon density is 
suggested by non-BBN contraints, and then compare the light 
element abundances which correspond to that density with those 
inferred from the observational data.  As an exercise of this 
sort, suppose (for reasons of ``naturalness" or inflation) that 
the Universe is ``flat": $\Omega_{\rm M} + \Omega_{\Lambda} 
= 1$.  When combined with the SN Ia magnitude-redshift data 
(Perlmutter et al. 1999), this suggests that $\Omega_{\rm M} 
= 0.29 \pm 0.07$ (and $\Omega_{\Lambda} = 0.71 \pm 0.07$).
Now, if this mass density estimate ($\Omega_{\rm M}$) is 
combined with with the X-ray determined cluster baryon fraction 
$f_{\rm B}$ (Evrard 1997; Steigman, Hata \& Felten 1999), 
the resulting nucleon abundance is $\eta_{10} = 4.5 \pm 1.5$.  
Although the uncertainty is large, it is reassuring that this 
non-BBN estimate has significant overlap with our SBBN estimate.
Indeed, for the baryon density in this range SBBN predicts:
(D/H)$_{\rm P} = 4.3 \pm 2.3 \times 10^{-5}$ and Y$_{\rm P} 
= 0.245 \pm 0.004$. 

\subsection{What SBBN May Do For Particle Physics}

The expansion rate of the early Universe is controlled by the
density of the relativistic particles present.  In the standard
model at the time of BBN these are: photons, electron-positron
pairs (when T $ \ga m_{e}$) and three ``flavors" of neutrinos
($\nu_{e}$, $\nu_{\mu}$, $\nu_{\tau}$) which, if ``light" ($m_{\nu}
\ll 1$~MeV), are relativistic at BBN even if one or more of them
may contribute to the present density of non-relativistic (``hot")
dark matter.  If ``new" particles were to contribute to the energy 
density at BBN, the increase in the density would result in an
increase in the universal expansion rate, leaving less time for
neutrons to transform into protons.  The higher $n/p$ ratio at BBN
would result in the production of more primordial \4he (Steigman,
Schramm \& Gunn 1977).  It is convenient (and conventional) to
characterize such additional contributions to the energy density
by comparing their effects to that of an additional ``flavor" of
(light) neutrino: $\Delta\rho \equiv \Delta N_{\nu}\rho_{\nu}$.
For $\Delta N_{\nu}$ small, $\Delta $Y $ \approx 0.01\Delta N_{\nu}$.
Notice in Figure 1 that the predicted \4he abundance is a little
high for perfect overlap with the observations.  If $\Delta N_{\nu}$ 
were $ < 0$, (N$_{\nu} \approx 2.8$) the overlap would improve (e.g., 
Hata et al. 1995), while if $\Delta N_{\nu} > 0$, the overlap would 
be reduced until it disappeared.  This is illustrated in Figure 2 
which shows the Y versus D/H BBN band that would result if $\Delta 
N_{\nu} = 0.2$ (i.e., $N_{\nu} = 3.2$, in contrast to the SBBN value 
of 3.0).  Notice that due to the faster expansion, more deuterium 
survives being burnt away so that, for fixed $\eta$, the D-abundance 
also increases; however since D/H is a much more sensitive function 
of $\eta$, this has a much smaller effect on the Y versus D/H relation 
than does the increase in Y.

\begin{figure}
\vspace{1.in}
\plotfiddle{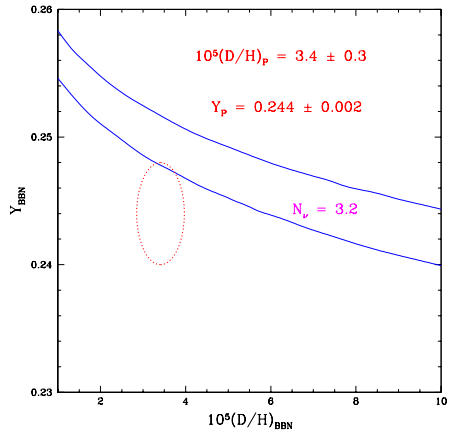}{1in}{0}{30}{30}{-100}{-50}
\caption{As Figure 1, but for $N_{\nu} = 3.2$}
\end{figure}

\section{Conclusions And Outlook}

The study of the early evolution of the Universe and, in
particular primordial nucleosynthesis, has truly entered
the precision era of cosmology.  Precise abundances of the 
light nuclides are predicted and inferred from observations
and the two are -- apparently -- in excellent agreement.
As pleased as we may be at this success, it behooves us
to avoid the temptation to rest on our laurels and to
test this consistency ever more carefully.  To this end,
it doesn't take much contemplation to identify several
clouds looming on the horizon.  What follows is my
personal list of some problems/issues I would like 
to see addressed at this Symposium and beyond.

\subsection{Problems/Issues}

First consider deuterium.  On the one hand, any determination
of the D/H ratio anywhere, anytime provides a {\it lower}
bound to the primordial abundance.  On the other hand, since
``wrong" velocity hydrogen can masquerade as deuterium, any
observation of ``deuterium" is really an {\it upper} bound 
to its true abundance.  More data tracking the velocity
structure of the absorbing features used to identify D 
and H and exploring variations in D/H in material with 
similar histories will be very valuable.  More data at 
high-redshift and low-metallicity will be very valuable.  
After all, at present we are drawing profound conclusions 
on the basis of only two such systems.

Much remains to be done concerning the primordial abundance
of \4he.  For the most part, the \hii regions from which
the helium abundance is inferred have been modelled as
homogeneous spheres or plane-parallel slabs.  A glance at 
the beautiful HST images of real \hii regions reveals that 
they are anything but such idealizations.  What are the effects 
of temperature and/or density inhomogeneities, and how large 
may they be?  What of underlying stellar absorption which, if 
present but neglected, would lead to an {\it under}estimate 
of the helium abundance.  And, what of the usually neglected
ionization correction for neutral hydrogen and helium (Viegas, 
Gruenwald \& Steigman 1999; see Viegas \& Gruenwald, this 
volume)?  Considering this latter work, where models of \hii 
regions ionized by realistic spectra of young star clusters 
were used in a reanalysis of the IT data, a {\it reduction} 
in \Yp of order 0.003 was suggested.  A comparison with Figure 
1 shows that if \Yp were reduced by this amount, the overlap 
between theory and data would, in fact, disappear.

\subsection{Wish List}

Given the setting of this Symposium (Natal) and the proximity
to the Christmas season, I'd like to conclude with my personal
wish list.  To avoid being greedy, I'll only ask for two gifts.

A half-dozen or so observations of deuterium in high-z, low-Z 
systems along the lines-of-sight to distant quasars, with D/H 
determined in each (on average) to 10\% or better.  With such 
a gift, I could determine $\eta$ to better than 4\%, predict
\Yp to $\la 0.0007$, and constrain $\Delta N_{\nu}$ to an 
uncertainty less than $\pm$~0.05.  I'd be a very happy 
cosmologist indeed.

My second wish is for \4he measured to 3\% accuracy (or better)
in each of about a dozen low-metallicity, extragalactic \hii 
regions, with care taken to address the several problems outlined 
above.  With such data, \Yp could be fixed to better than the 
current level of $\pm 0.002$, permitting \4he to be used as a 
baryometer ($\Delta\eta/\eta~\la 20\%$).


\acknowledgments

Much of what I know about this subject I have learned from my 
collaborators and I would be remiss if I failed to thank them 
for their contributions.  In particular, I wish to acknowledge
R. Gruenwald, K. Olive, E. Skillman, M. Tosi, and S. M. Viegas 
and, of course, my late friend Dave Schramm.  L. da Silva, M. 
Spite and the Scientific and Local Organizing committees deserve 
great credit for their efficient organization of a very enjoyable 
and successful meeting.  In part, this work is supported at The 
Ohio State University by DOE grant DE--AC02--76ER--01545.

\end{document}